\documentstyle[epsfig,twocolumn]{article}
\title{ \vspace{-2.cm}
  \hspace*{11cm} {\large Budker INP Preprint 99-65} \\
  \hspace*{13.5cm} {\large July 1999} \\
  \vspace{0.3cm} {\LARGE\bf Status of gamma-gamma, gamma-electron
    colliders } \thanks{Talk at International Conference on the
    Structure and Interactions of the Photon (Photon 99), Freiburg,
    Germany, 23-27 May 1999, Submitted to Nucl. Phys. Proc. Suppl. B}}

\author{Valery Telnov \\
  Institute of Nuclear Physics, 630090 Novosibirsk,
  Russia \thanks{email:telnov@inp.nsk.su}} 

\date{}

\begin{document}
\newcommand{\M}{\mbox{m}}
\newcommand{\n}{\mbox{$n_f$}}
\newcommand{\EP}{\mbox{e$^+$}}
\newcommand{\EM}{\mbox{e$^-$}}
\newcommand{\EPEM}{\mbox{e$^+$e$^-$}}
\newcommand{\EMEM}{\mbox{e$^-$e$^-$}}
\newcommand{\GG}{\mbox{$\gamma\gamma$}}
\newcommand{\GE}{\mbox{$\gamma$e}}
\newcommand{\GP}{\mbox{$\gamma$e$^+$}}
\newcommand{\TEV}{\mbox{TeV}}
\newcommand{\GEV}{\mbox{GeV}}
\newcommand{\LGG}{\mbox{$L_{\gamma\gamma}$}}
\newcommand{\LGE}{\mbox{$L_{\gamma e}$}}
\newcommand{\LEE}{\mbox{$L_{ee}$}}
\newcommand{\WGG}{\mbox{$W_{\gamma\gamma}$}}
\newcommand{\EV}{\mbox{eV}}
\newcommand{\CM}{\mbox{cm}}
\newcommand{\MM}{\mbox{mm}}
\newcommand{\NM}{\mbox{nm}}
\newcommand{\MKM}{\mbox{$\mu$m}}
\newcommand{\SEC}{\mbox{s}}
\newcommand{\CMS}{\mbox{cm$^{-2}$s$^{-1}$}}
\newcommand{\MRAD}{\mbox{mrad}}
\newcommand{\IND}{\hspace*{\parindent}}
\newcommand{\E}{\mbox{$\epsilon$}}
\newcommand{\EN}{\mbox{$\epsilon_n$}}
\newcommand{\EI}{\mbox{$\epsilon_i$}}
\newcommand{\ENI}{\mbox{$\epsilon_{ni}$}}
\newcommand{\ENX}{\mbox{$\epsilon_{nx}$}}
\newcommand{\ENY}{\mbox{$\epsilon_{ny}$}}
\newcommand{\EX}{\mbox{$\epsilon_x$}}
\newcommand{\EY}{\mbox{$\epsilon_y$}}
\newcommand{\BI}{\mbox{$\beta_i$}}
\newcommand{\BX}{\mbox{$\beta_x$}}
\newcommand{\BY}{\mbox{$\beta_y$}}
\newcommand{\SX}{\mbox{$\sigma_x$}}
\newcommand{\SY}{\mbox{$\sigma_y$}}
\newcommand{\SZ}{\mbox{$\sigma_z$}}
\newcommand{\SI}{\mbox{$\sigma_i$}}
\newcommand{\SIP}{\mbox{$\sigma_i^{\prime}$}}
\maketitle

\begin{abstract}

  This report on Photon Colliders briefly reviews three main issues:
physics motivation, possible parameters and technical feasibility,
plans of works and international cooperation. New scheme of laser
optics at the interaction region is described which can drastically
(at least by one order) reduce the cost of the laser system.

\end{abstract}
\section {Introduction}
 As you certainly know, Linear colliders in the range of a few hundred
GeV to 1.5 TeV range are under intense study around the world.  Three
specific project studies in Europe, Asia, and North America are going
forward, with the intent to submit full conceptual design reports in
the 2001-2002 time frame.  In parallel, several hundred high energy
physicists are contributing to advancing the physics case for linear
colliders, and optimizing detector design and technologies.

In addition to \EPEM\ collisions, linear colliders provide a unique
possibility to study \GG\ and \GE\ interactions at energies and
luminosities comparable to those in \EPEM\
collisions~\cite{GKST81}-\cite{PToday}.  High energy
photons for \GG, \GE\ collisions can be obtained using laser
backscattering.  Modern laser technology presents the real possibility
for construction of the laser system  for \GG, \GE\ collider
('photon collider').  This option is included now in the
pre-conceptual design of the NLC (North American)~\cite{NLC}, TESLA
(European)~\cite{TESLA} and JLC (Asian)~\cite{JLC} projects, and is
attracting an increasing interest among both theorists and
experimentalists.

  However, in our time of tight HEP budgets the physics community
needs a very clear answer to the following question: a) can \GG,\GE\
collisions give new physics information in addition to \EPEM\
collisions that could justify an additional collider cost ($\sim$15\%,
including detector); b) is it technically feasible; c) is there enough
people who are ready to spend a significant part of their career
for the design and construction of a photon collider, and
exploiting its unique science?  

Shortly, my  answers are the following: 

a) Certainly yes. There are many predictions of extremely interesting
physics in the region of the next linear colliders. If something new will
be discovered (Higgs, supersymmetry or ... quantum gravity with extra
dimensions), to understand better a nature of these new
phenomena they should be studied in different reactions which give
complementary information.

b) There are no show-stoppers. There are good ideas on obtaining 
very high luminosities, on laser and optical schemes. It is clear how to
remove disrupted beams and there is an understanding of backgrounds.
However, much remains to be done in terms of detailed studies and
experimental tests.  Special efforts are required for the development
of the laser and optics which are the key elements of photon
colliders.  

c) This is a new direction, not well known to physics community, and,
as usually, it has to pass several natural phases of development.  In the
last decade, there has been  growing  interest to \GG, \GE\
collisions and the bibliography of recent reports on \GG, \GE\ physics
now numbers over 1000 papers, mostly theoretical. The next phase will
require  much wider participation of the experimental community. 

 To this end, recently, it was decided to initiate International
collaboration on Photon Colliders.  This Collaboration does not
replace the regional working groups, but rather supports and
strengthens it. The Invitation letter, signed by Worldwide Study
contact persons on photon colliders: V.Telnov (Europe), K. Van Bibber
(North America), T.Takahashi (Asia) will be send to you shortly. 
\section{Physics}
    The most interesting physics (``expected'' discoveries) at next
linear colliders is the search for and study of the Higgs boson(s),
supersymmetric particles, and many other new phenomena such as quantum
gravity (very popular topic in the last year).  Photon colliders can
make a considerable contribution.
\subsection{Higgs}
\begin{figure}[!htb]
\centering
\vspace*{-.7cm} 
\hspace*{-0.8cm} \epsfig{file=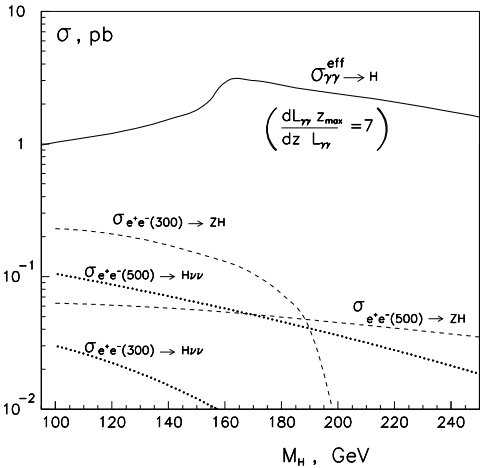,width=9.cm,angle=0} 
\vspace*{-1.3cm} 
\caption{Cross sections for the Standard model Higgs in \GG\ and
 \EPEM\ collisions.}
\vspace{1mm}
\label{cross}
\end{figure}

  The Higgs boson (which is thought to be responsible for the origin
of particle masses) will be produced at photon colliders as a single
resonance.  The cross section is proportional to the two-photon decay
width of the Higgs boson which is very sensitive to all heavy charged
particles (even super-heavy) which get their mass via the Higgs
mechanism. The mass of the Higgs most probably lies in the region of
100$<M_H<$250 GeV. The effective cross section is presented in
Fig.~\ref{cross}~\cite{ee97}.  
Note that here \LGG\ is defined as the \GG\ luminosity at the high
energy luminosity peak ($z=\WGG/2E_e>0.65$ for $x=4.8$) with FWHM
about 15\%. For comparison, the cross sections of the Higgs production
in \EPEM\ collisions are shown in the same figure.

We see that for $M_H=$ 120--250 GeV the effective cross section in
\GG\ collisions is larger than that in \EPEM\ collisions by a factor
of about 6--30!  If the Higgs is light enough, its width is much less
than the energy spread in \GG\ collisions. It can be detected as a
peak in the invariant mass distribution or can be searched by energy
scanning using the very sharp ($\sim 1\%$) high energy edge of luminosity
distribution~\cite{ee97}.

  Observation of a sharp step in the visible cross section will imply
narrow resonance production with subsequent decay in the considered
channel. This method is very attractive for the study of the Higgs in the
$\tau\tau$ decay mode where the direct reconstruction is impossible due to
undetected neutrinos while it can be seen as a step in visible cross
section for events consisting of two low multiplicity collinear jets.
The total number of events in the main decay channels $H \to b\bar b,
WW(W^*), ZZ(Z^*)$ will be several thousands for a typical integrated
luminosity of 10 fb$^{-1}$. The scanning method also enables
the measurement of the Higgs mass with a high precision.
\subsection{Charge pair production}
   The second example is the charged pair production. It could be
   $W^+W^-$ or $t\bar{t}$ pairs or some new, for instance,
   supersymmetric particles.  Cross sections for the production of
   charged scalar, lepton, and top pairs in \GG\ collisions are larger
   than those in \EPEM\ collisions by a factor of approximately 5--10;
   for WW production this factor is even larger, about 10--20. The
   corresponding graphs can be found
   elsewhere~\cite{TEL90},\cite{TESLA},\cite{ee97}.

\begin{figure}[!thb]
\centering
\vspace*{-0.9cm}
\hspace*{-0.5cm} \epsfig{file=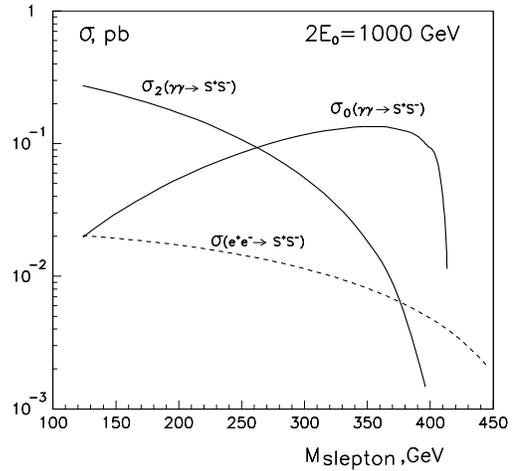,width=9.2cm}
\vspace*{-1.3cm}
\caption{ Cross sections for charged scalars production in \EPEM\ and
\GG\ collisions at $2E_0$ = 1 TeV collider (in \GG\ collision
$W_{max}\approx 0.82$ GeV, $x=4.6$); $\sigma_0$ and $\sigma_2$ correspond to
the total \GG\ helicity 0 and 2.}
\vspace{1mm}
\label{crossel}
\end{figure}

 The cross section of the scalar pair production
(sleptons, for example) in collision of polarized photons is shown
in Fig.\ref{crossel}.
One can see that for heavy scalars the cross section in collisions of
polarized photons is higher than that in \EPEM\ collisions by a factor
of 10--20. The cross section near the threshold is very sharp (in
\EPEM\ it contains a factor $\beta^3$) that can be used for
measurement of particle masses. 

Note that for scalar selectrons the cross section in \EPEM\ collisions
is not described by the curve in Fig.\ref{crossel} due to existence of
additional exchange diagram (exchange by neutralino), correspondingly
the cross section is not described by pure QED (as it takes place in
\GG). Measurement of cross sections in both channels
give, certainly, complimentary information.
\subsection{Accessible masses}
 In \GE\ collisions, charged supersymmetric particles with masses
  higher than those in \EPEM\ collisions can be produced (a heavy charged
  particle plus a light neutral); \GG\ collisions also provide higher
  accessible masses for particles which are produced as a single
  resonance in \GG\ collisions (such as the Higgs boson). 
This is very important argument.

One very close example. It is very likely that LEP-II does not see the
Higgs because its energy is only somewhat lower than $\EPEM\ \to ZH$
threshold. Having \GG\ mode at LEP-II (this is impossible for storage
rings, of course) one could produce Higgs with the mass higher on 60
GeV.  The same story is with the search for the supersymmetry where
\GE\ mode could help very much.
\subsection{Quantum gravity effects in Extra Dimensions.}
  This new theory~\cite{Arkani} is very interesting though beyond my
imagination. It suggests one of possible explanation why gravitation
forces are so weak in comparison with electroweak forces.

According to this theory the gravitational forces are as strong as
electroweak forces at small distances in space with extra dimensions
and became weak at large distances due to ``compactification'' of these
extra dimensions. 

  It turns out that this extravagant theory can be tested at linear
colliders and according to T.Rizzo~\cite{RIZZO} ($\GG\ \to WW$) and
K.Cheung~\cite{CHEUNG} ($\GG \to \GG$) photon colliders are sensitive
to a factor of 2 higher quantum gravity mass scale than \EPEM\ collisions.
\section{Luminosity of photon colliders in current designs.}
\subsection{0.5--1 TeV colliders}
    Some results of simulation of \GG\ collisions at TESLA, ILC
(converged NLC and JLC) and CLIC are presented below. Beam parameters were
taken the same as those in \EPEM\ collisions with the exception of
horizontal beta function at IP, which is taken  equal to
2 mm for all cases. In \GG\ collisions, the beamstrahlung is absent and
the horizontal size can be made much smaller than that in \EPEM\
collisions. Minimum $\beta_x$ is determined by the Oide effect
(radiation in quads) which is included in the simulation code and also
by technical problems connected with the chromatic corrections in both
transverse directions -- the limit here is not clear so far. The
conversion point(CP) is situated at  distance
$b=\gamma\sigma_y$. It is assumed that electron beams have 85\%
longitudinal polarization and laser photons have 100\% circular
polarization.

% The simulation code~\cite{TEL95} takes into account all important
%processes: linear Compton scattering with all polarization effects,
%beamstrahlung (without polarization effects), coherent pair creation and
%interaction between charged particles.

\vspace{-0.3cm}
{\setlength{\tabcolsep}{0.1mm}
{\footnotesize
\begin{table}[!hbtp]
\caption{Parameters of  \GG\ colliders based on Tesla(T), ILC(I)
and CLIC(C).}
\vspace{-0.2cm}
\begin{center}
\hspace*{-2.3mm}\begin{tabular}{c c c c c c c} \hline
 & T(500) & I(500) & C(500) \ &
  T(800) & I(1000) &  C(1000) 
                                                   \\ \hline \hline 
\multicolumn{7}{c}{ no deflection, $b=\gamma \sigma_y$, $x=4.6$} \\ \hline
$N/10^{10}$& 2. & 0.95 & 0.4 & 1.4 & 0.95 & 0.4 \\  
$\sigma_{z}$, mm& 0.4 & 0.12 & 0.05 & 0.3 & 0.12 & 0.05 \\  
$f_{rep}\times n_b$, kHz& 15 & 11.4 &30.1& 13.5 & 11.4 & 26.6 \\
% $\Delta t_b$, ns  & 708 & 354 & 708 & 708 & 6 & 6 \\ 
$\gamma \epsilon_{x,y}/10^{-6}$,m$\cdot$rad & $10/0.03$ & $5/0.1$ & 
$1.9/0.1$&  $8/0.01$ & $5/0.1$ & $1.5/0.1$ \\
$\beta_{x,y}$,mm at IP& $2/0.4$ & $2/0.12$ & $2/0.1$ &
$2/0.3$& $2/0.16$ & $2/0.1$ \\
$\sigma_{x,y}$,nm& $200/5$ & $140/5$ & $88/4.5$ & 
$140/2$ & $100/4$ & $55/3.2$ \\  
b, mm & 2.4 & 2.4 & 2.2 & 1.5 & 4 & 3.1 \\
$L(geom),\,\,\,  10^{33}$& 48 & 12 & 10 & 75 & 20 & 19.5\\  
$\LGG (z>0.65), 10^{33} $ & 4.5 & 1.1 & 1.05 & 7.2 & 1.75 & 1.8 \\
$\LGE (z>0.65), 10^{33}$ & 6.6 & 2.6 & 2.8 & 8  & 4.2 & 4.6 \\
$\LEE, 10^{33}$ & 1.2 & 1.2 & 1.6 & 1.1 & 1.8 & 2.3 \\
$\theta_x/\theta{_y},_{max}$, mrad ~ & 5.8/6.5 & 6.5/6.9 & 6/7& 
 4.6/5 & 4.6/5.3 & 4.6/5.5 \\ \hline
\vspace{-5.mm}
\end{tabular}
\end{center}
\label{table1}
\end{table}
}} 

We see that \GG\ luminosity in the hard part of the spectrum is $\LGG
  (z>0.65)\sim 0.1L(geom)\sim (1/6)L_{\EPEM}$.  Beside \GG\
  collisions, there is considerable \GE\ luminosity and it is possible
  to study \GE\ interactions simultaneously with \GG\
  collisions. 
  
\begin{figure}[!htb]
\centering
\vspace*{-1.8cm} 
\hspace*{-1.4cm} \epsfig{file=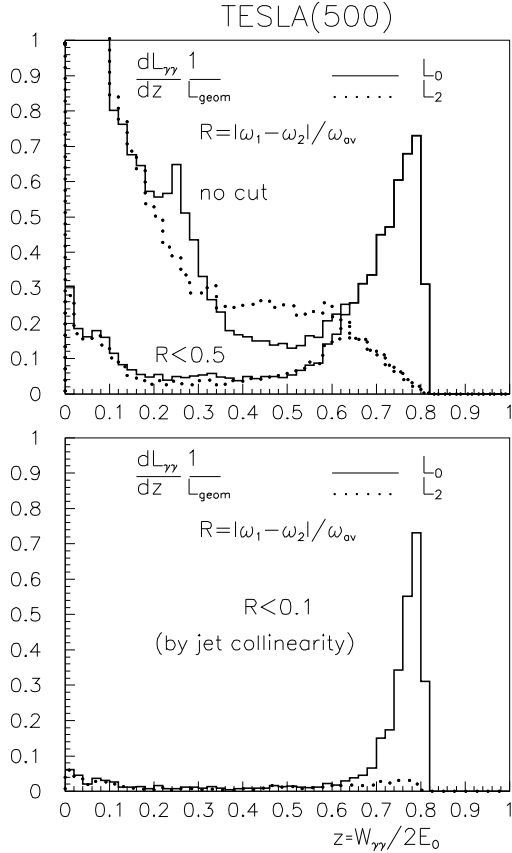,width=9.8cm,angle=0} 
\vspace*{-1.9cm} 
\caption{\GG\ luminosity spectra at TESLA(500) for parameters
presented in Table 1. Solid line for total helicity of two photons 0
and dotted line for total helicity 2. Upper curves without cuts, two
lower pairs of curves  have cut on the relative difference
of the photon energy. See comments in the text.}
\vspace*{-2mm} 
\label{TeslaR}
\end{figure}

   The normalized \GG\ luminosity spectra for a 0.5 TeV TESLA are
   shown in Fig.\ref{TeslaR}(upper).
The luminosity spectrum is decomposed into two parts, with the total
helicity of two photons 0 and 2. We see that in the high energy part
of the luminosity spectra photons have high degree of polarization,
which is very important for many experiments.  In addition to the high
energy peak, there is a factor 5--8 larger low energy luminosity. It
is produced by photons after multiple Compton scattering and
beamstrahlung photons. Fortunately, these events have large boost and
can be easily distinguished from the central high energy events.  In
the same Fig.\ref{TeslaR}(upper) you can see the same spectrum with an
additional ``soft'' cut on the longitudinal momentum of the produced
system which suppresses low energy luminosity to a negligible level.

Fig.\ref{TeslaR} (lower) shows the same spectrum with a stronger cut
on the longitudinal momentum. In this case, the spectrum has a nice
peak with FWHM about 7.5\%. Of course, such procedure is somewhat
artificial because instead of such cuts one can directly selects
events with high invariant masses, the minimum width of the invariant
mass distribution depends only on the detector resolution. However,
there are very important examples when one can obtain a ``collider
resolution'' somewhat better than the detector resolution, such as the
case of only two jets in the event when one can restrict the
longitudinal momentum of the produced system using the acollinearity
angle between jets ($H\to b\bar b, \tau\tau$, for example).
\subsection{\GG\ collider for low mass Higgs}
  It is very possible that the Higgs boson has a mass in the region
115-150 GeV as predicted in some theories. It is of interest to
consider possible parameters of a \GG\ collider based on TESLA and ILC
at these energies. Two variants were considered for H(130): 1) the
``Compton'' parameter $x$ is fixed near the threshold of \EPEM\
creation ($x \approx 4.6$), which corresponds to $\lambda \sim 325$ nm
and $E_0=79$ GeV; 2) the laser is the same as for $2E_0 = 500$ GeV
colliders, namely a Nd:glass laser with $\lambda=1.06\;\mu m$, which
corresponds to $x=1.8$ and $E_0=100$ GeV. All other beam parameters
are taken the same as for $2E_0 = 500$ GeV (see
Table~\ref{table1}). Results of simulation for these two cases are
shown in Table \ref{table2} (TESLA and ILC) and in Fig.\ref{Higgs130}
(TESLA).  Comparing these two variants we can conclude that one can
use the same Nd:glass laser at all energies below $2E_0 \sim$ 500 GeV.

{\setlength{\tabcolsep}{1mm}
{\footnotesize
\begin{table}[!hbt]
\vspace{0.cm}
\caption{Parameters of the \GG\ colliders for 
Higgs(130) at TESLA(T)and ILC(I).}
\vspace{-0.0cm}
\begin{center}
\hspace*{-2mm}\begin{tabular}{c c c  c c} \hline
 & T(2x100) & I(2x100) & T(2x79) & I(2x79) 
                                                   \\ \hline \hline 
%\multicolumn{5}{c}{ no deflection, $b=\gamma \sigma_y$} \\ \hline
\multicolumn{5}{c}{\mbox{\hspace*{2.6cm}} $x=1.8$ \mbox{\hspace*{1.4cm}}\
 $x=4.6$ } \\ \hline
$\sigma_{x,y}$,nm& $320/7.8$ & $230/7.8$ & $360/8.8$ & $250/8.8$ \\  
b, mm & 1.5 & 1.5 & 1.4 & 1.4 \\
$L(geom),\,\,\,  10^{33}$& 19 & 4.6 &  15 & 3.7 \\  
$\LGG (z/z_m>0.8),10^{33}$ & 1.55 & 0.37 & 1.45 & 0.35 \\
$\LGE (z/z_m>0.8),10^{33}$ & 3. & 1.45 & 1.7 & 0.83 \\
$\theta_x/\theta{_y},_{max}$, mrad ~ & 5.2/6.2 & 5.2/7 & $\sim$ 10/12 & 
 $\sim$ 10/12 \\ \hline
\end{tabular}
\end{center}
\label{table2}
\vspace{-8mm}
\end{table}
} 
\begin{figure}[!h]
\centering
\vspace*{-1.2cm} 
\hspace*{-1.5cm} \epsfig{file=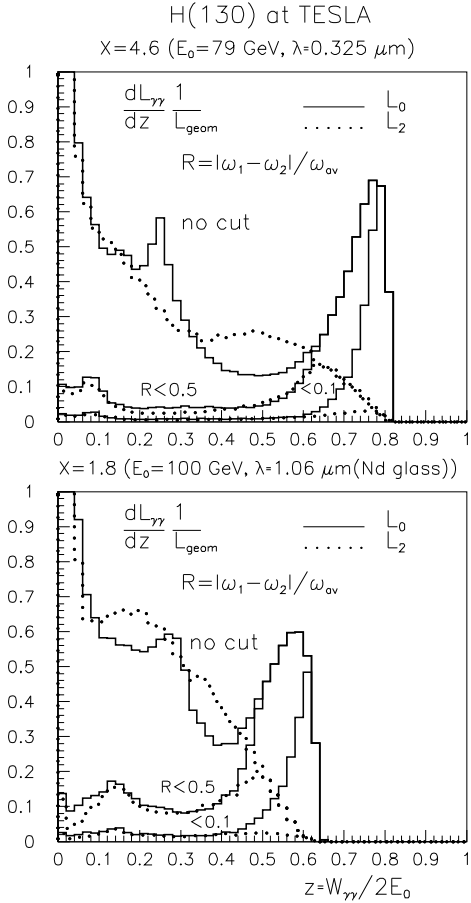,width=9.3cm,angle=0} 
\vspace*{-2.2cm} 
\caption{Luminosity spectra of \GG\ collision of ``low'' energy \GG\ collider
(TESLA beam parameters) for study of the Higgs with a mass
$M_H=130$ GeV, upper figure for $x=4.8$ and lower for $x=1.8$
(the same laser as for $2E_0=500$ GeV).}
\vspace{-0.3cm}
\label{Higgs130}
\end{figure} 
\section{Ultimate \GG, \GE\  luminosities }
   The \GG\ luminosities in the current projects are determined by the
``geometric'' luminosity of the electron beams. The only collision
effect restricting the maximum value of the \GG\ luminosity is the
coherent pair creation when the high energy photon is converted into
an \EPEM\ pair in the field of the opposing electron
beam~\cite{CHEN},\cite{TEL90}.  Having electron beams with smaller
emittances one can obtain much higher \GG\
luminosity~\cite{TSB2}. Fig.\ref{sigmax} shows dependence of the \GG\
(solid curves) and \GE\ (dashed curves) luminosities on the horizontal
beam size.
\begin{figure}[!htb]
\centering
\vspace*{-1.2cm} 
\hspace*{-0.8cm} \epsfig{file=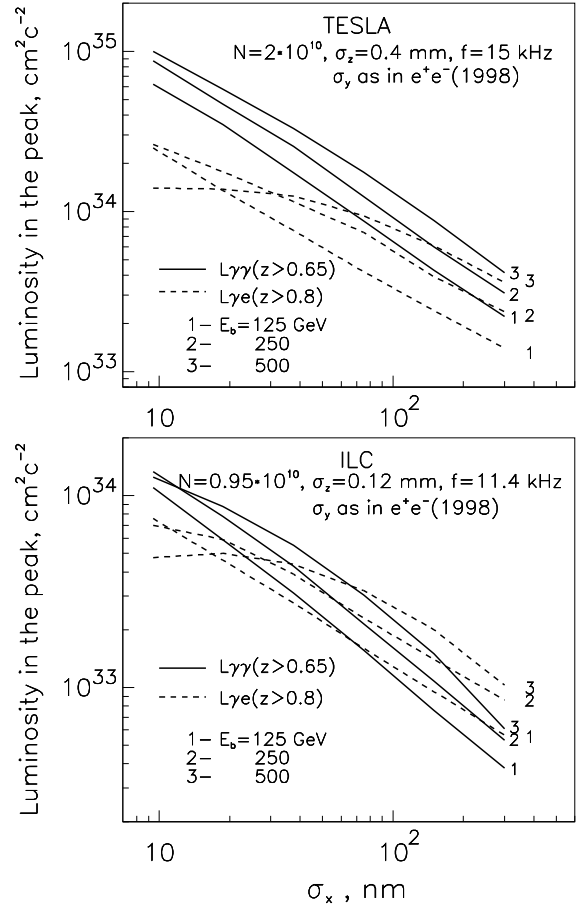,width=9.cm,angle=0} 
\vspace*{-1.8cm} 
\caption{Dependence of \GG\ and \GE\ luminosities in the high energy
peak on the horizontal beam size for TESLA and ILC at various
energies. See also comments in the text.}
\vspace{0mm}
\label{sigmax}
\vspace{-2mm}
\end{figure} 
 The vertical emittance is taken as in TESLA(500), ILC(500)
projects (see Table \ref{table1}). The horizontal beam size was varied
by change of horizontal beam emittance keeping the horizontal beta
function at the IP constant and equal 2 mm.

One can see that all curves for \GG\ luminosity follow their natural
behavior: $\L\propto 1/\sigma_x$, with the exception of ILC at
$2E_0=1$ GeV where at small $\sigma_x$ the effect of coherent pair
creation is seen.\footnote{This curve has also some bend at largest
$\sigma_x$ that is connected with synchrotron radiation in quads (Oide
effect) due to a large horizontal emittance. One can avoid this effect
by taking larger $\beta_x$ and smaller \ENX.}  This means that at the
same colliders the \GG\ luminosity can be increased by decreasing
horizontal beam size at least by one order ($\sigma_x < 10$ nm is
difficult due to some effects connected with crab
crossing). Additional increase of \GG\ luminosity by a factor about 3
(TESLA), 7(ILC) can be obtained by further decrease of the vertical
emittance~\cite{TKEK}. So, if to use beams with smaller emittances, the
\GG\ luminosity at TELSA, ILC can be increase by almost 2 orders of
magnitude. However, even with one order improvement, the number of
``interesting'' events (the Higgs, charged pairs) at photon colliders
will be larger than that in \EPEM\ collisions by about one order. This
is a nice goal and motivation for photon colliders.

  In \GE\ collision (Fig.\ref{sigmax}, dashed curves), the behavior of
the luminosity on $\sigma_x$ is different due to additional collisions
effects: beams repulsion and beamstrahlung. As a result, the
luminosity in high energy peak is not proportional to the
``geometric''  luminosity.

   There are several ways of decreasing transverse beam emittances
(their product): optimization of storage rings with long wigglers,
development of low-emittance RF or pulsed photo-guns with merging many
beams with low emittances.  Here some progress is certainly
possible. Moreover, there is one method which allows further decrease
of beam cross sections by two orders in comparison with current
designs. It is laser cooling~\cite{TSB1},\cite{Monter}.  

In the method of laser cooling the electron beam at an energy of
several GeV is collided 1--2 times with a powerful laser flash, losing
in each collision a large fraction ($\sim 90\%$) of its energy to
radiation, with reacceleration between cooling sections. The physics
of the cooling process is the same as in a wiggler.  One of problems
here is the required laser flash energy, it should be about 10-100 J
depending on beam energy, laser wave length and optical scheme. One
very promising variant of laser optics for laser cooling is discussed
in the next section. Other problem here is capture (equal to focusing)
of the electron beam with the large energy spread (about 10-15 \% at
$E \sim 0.5$ GeV) without dilution of the emittance. The similar
problem has been solved for final focusing of beams at linear
colliders, where it is also necessary to correct effects of
chromaticity to high orders. The corresponding parameter of the
problem $(F/\beta)\times (\sigma_E/E)$ in the laser cooling is smaller
and the energy is smaller, so there are hopes that such magnetic
system can be build.
\section{New ideas on laser optics.}
   The laser flash energy required for conversion of 65\% (one
collision length) of electrons to high energy photons is about
1.5(2.5) J for ILC (TESLA). At collision rate 10-15 kHz, the average
laser power will be about 20-30 kW. Such system will be a huge and
expensive. Livermore experts give cost estimate to such laser system of the
order of 200 M\$~\cite{PERRY}.  

Fortunately, there is a solution which can decrease the cost by one
order (at least). One flash contains about $10^{19}$ laser photons and
only $10^{10}$ photons are knocked out in collision with one electron
bunch. It is very natural to use laser pulse many times, and optics
presents us such a possibility. Shortly, the method is the
following. Using the train of low energy pulses from the laser one can
create in the external passive cavity (with one mirror having some
small transparency) an optical pulse of the same duration but with
much higher energy (pulse stacking). This pulse circulates many times in
the cavity each time colliding with electron bunches passing the
center of the cavity. 

The idea of pulse stacking is simple but not trivial and not well
known in HEP community (and even to laser experts, though it is as old
as the Fabry-Perot interferometer). This method is used now in several
experiments on detection of gravitation waves. It was mentioned also
in NLC ZDR~\cite{NLC} though without analysis and further development.

   To my opinion, pulse stacking is very natural for photon colliders
and allows not only to build relatively cheap laser system for
$e\to\gamma$ conversion but give us the practical way for realization of
laser cooling, i.e. opens up the way to ultimate luminosities of photon
colliders. 

As it is the key problem of photon colliders, let me consider this
method in more detail. The principle of pulse stacking  is shown in
Fig.\ref{cavity}.
\begin{figure}[!htb]
\centering
\vspace*{0.2cm} 
\hspace*{-0.2cm} \epsfig{file=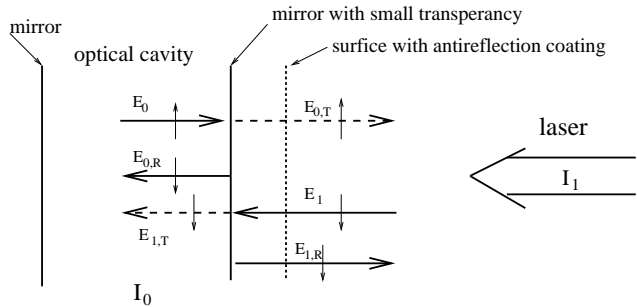,width=8.5cm,angle=0} 
\vspace*{-0.cm} 
\caption{Principle of pulse stacking in an external optical cavity.}
\vspace{2mm}
\label{cavity}
\vspace{-0mm}
\end{figure} 
The secret consists in the following. There is well known optical
theorem: at any surface, the reflection coefficients for light coming
from one and other sides have opposite signs. In our case, this means
that light from the laser entering through semi-transparent mirror into
the cavity interferes with reflected light inside the cavity {\bf
constructively}, while the light leaking from the cavity interferes
with the reflected laser light {\bf destructively}. Namely, this fact produces
asymmetry between cavity and space outside the cavity! 

Let R be the reflection coefficient, T is transparency coefficient
and $\delta$ are passive losses in the right mirror. From the energy
conservation $R+T+\delta =1$. Let $E_1$ and $E_0$ be the amplitudes
of the laser field and the field inside the cavity. In equilibrium,
\begin{equation}
E_0= E_{0,R} + E_{1,T}. 
\end{equation}
Taking into account that $E_{0,R}=E_0\sqrt{R}$, $E_{1,T}=E_1\sqrt{T}$ and 
$\sqrt{R}\sim 1-T/2-\delta/2$ for $R\approx 1$ we obtain
\begin{equation}
E_0^2= E_1^2\frac{4T}{(T+\delta)^2}. 
\end{equation} 
The maximum ratio of intensities is obtained at $T=\delta$, then 
\begin{equation}
I_0/I_1=1/\delta \approx Q,
\end{equation} 
where $Q$ is the quality factor of the optical cavity.  Even with two
metal mirrors inside cavity, one can hope to get the gain factor about
50-100, with multi-layer mirrors it could reach $10^5$. ILC(TESLA)
colliders have 120(2800) electron bunches in the train, so the factor
100(1000) would be perfect for our goal, but even the factor of ten
means the drastic reduction of the cost.

   Obtaining of high gains requires a very good stabilization of cavity
size: $\delta L \sim \lambda/4\pi Q$, laser wave length: $\delta
\lambda/\lambda \sim \lambda/4\pi QL$ and distance between the laser
and the cavity: $\delta s \sim\lambda/4\pi$. Otherwise, the  condition of
construction interference will be not fulfilled. Besides, the
frequency spectrum of the laser should coincide with the cavity modes,
that is automatically fulfilled when the ratio of the cavity length and
that of laser oscillator is equal to integer number 1, 2, 3... . 

For $\lambda = 1\;\mu m$ and $Q=100$, the stability of the cavity
length should be about $10^{-7}$ cm. In the LIGO experiment
 on detection of gravitational waves which uses 
similar techniques with $L\sim 4$ km and $Q\sim 10^5$ the expected
sensitivity is about $10^{-16}$ cm.  In comparison with this project
our goal seems to be very realistic.

      In HEP literature I have found only one reference on pulse
stacking of short pulses ($\sim 1$ ps) generated by FEL~\cite{HAAR}
with the wave length of 5 $\mu$m. They observed pulses in the cavity
with 70 times the energy of the incident FEL pulses, though no long
term stabilization was done.
   
     Possible layout of optics at the interaction region scheme is
shown in Fig.\ref{optics}. In this variant, there are two optical
cavities (one for each colliding electron beam) placed outside the
electron beams.

Another possible variant has only one cavity common for both electron
beams. In this case, it is also possible to arrange two conversion
points separated by the distance of several millimeters (as it is
required for photon colliders), though the distribution of the field
in the cavity is not completely stable in this case (though may be
sufficient for not too large Q). Also, mirrors should have holes for
electron beams (which does not change Q factor of the cavity too
much).

\begin{figure}[!htb]
\centering
\vspace*{-0.5cm} 
\hspace*{-0.4cm} \epsfig{file=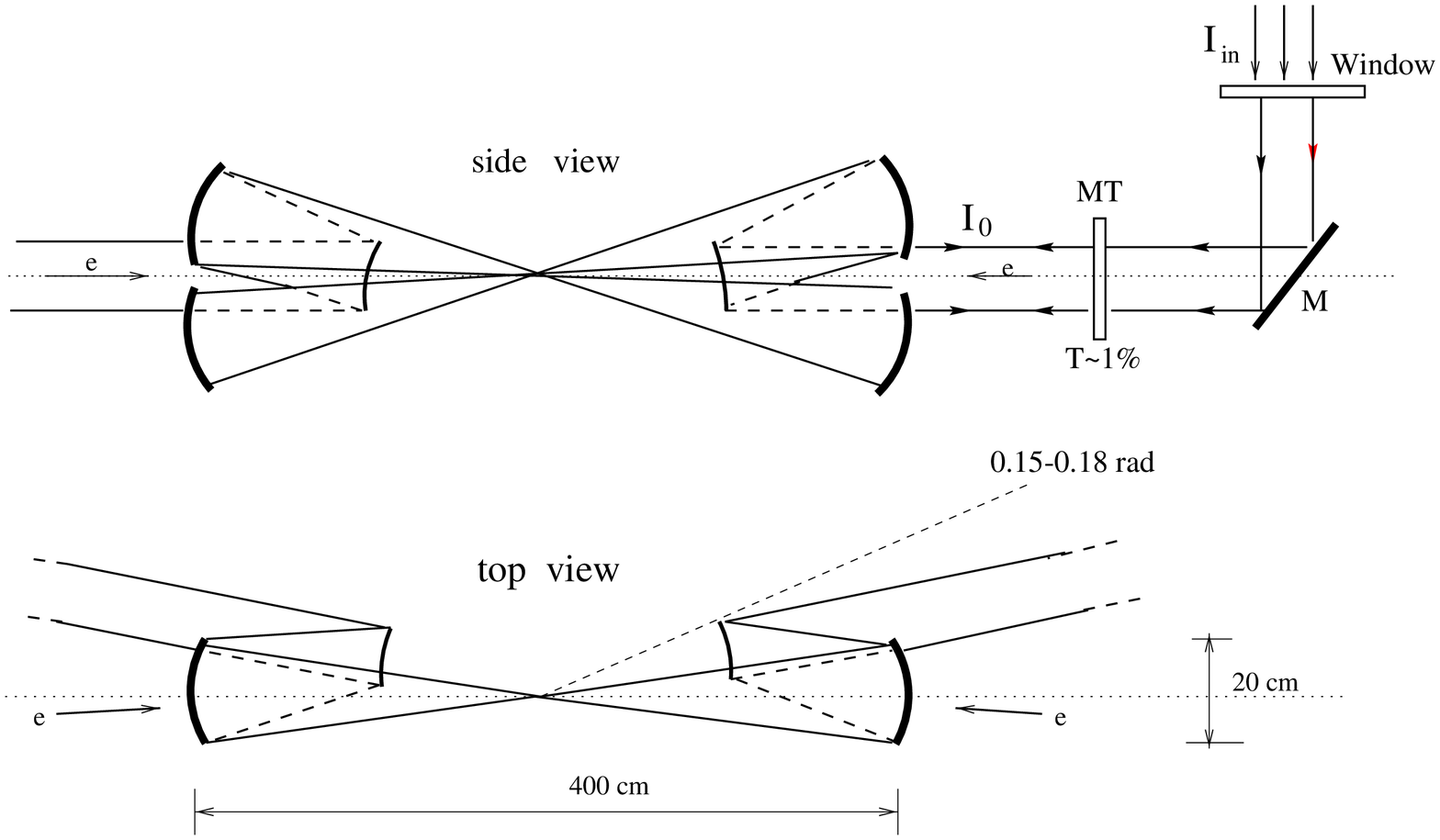,width=9cm,angle=0} 
\vspace*{-0.1cm} 
\caption{Possible scheme of optics at the IR.}
\vspace{0mm}
\label{optics}
\vspace{1mm}
\end{figure} 

  The use of the pulse stacking in the optical cavity make the
idea of laser cooling (previous section) very realistic, though the required
flash energy should be more than by one order higher than that required for
$e \to \gamma$ conversion.   
\section{Conclusion}
   Prospects of photon colliders for particle physics are great; the physics
community should not miss this unique possibility.     
\section{Acknowledgement}
I would like to thank K. van Bibber, A. de Roeck, A. Skrinsky,
T. Takahashi, M. Xie, K. Yokoya for useful discussions.

\end{document}